\documentclass{iopjournal}

\usepackage{amsmath}
\usepackage{fancyhdr}
\usepackage{graphicx}
\usepackage{hyperref}
\usepackage{xcolor}
\usepackage{ragged2e}
\justifying
\begin{document}

\articletype{Article type} 

\title{Electro-optically controlled photon group velocity, temporal walk-off and two-photon entanglement via nematic liquid crystal}

\author{Gyaprasad$^{1,*}$\orcid{0000-0003-3275-0052}and Rajneesh Joshi$^2$\orcid{0000-0001-5864-697X} }

\affil{$^1$Department of Physics, Government Degree College Jaithra, 207249 Etah, Uttar Pradesh, India}

\affil{$^2$Department of Physics, Government Degree College Danya, 263622 Almora, Uttarakhand, India}

\affil{$^*$Author to whom any correspondence should be addressed.}

\email{gyaprasadamu@gmail.com}


\begin{abstract}
\justifying
The propagation of the quantum states of light in dispersive and anisotropic media is a fundamental problem in quantum optics. We present a unified theoretical framework for the propagation of the quantum states of light in voltage-controlled nematic liquid crystals, incorporating both material dispersion and electrically tunable birefringence. By treating photons as finite-bandwidth wave packets, we derive analytical expressions for group velocity, temporal walk-off, and phase evolution of orthogonally polarized modes. The results demonstrate that nematic liquid crystals can serve as electrically tunable quantum photonic devices capable of manipulating photon arrival times, polarization correlations, and temporal indistinguishability of entangled photon pairs. These results show the direct relevance to quantum communication and photonic quantum information processing.
\end{abstract}

\section{Introduction}
The propagation of light pulses through dispersive media has long been studied in classical optics\cite{born2013principles}, in which the concept of group velocity describes the motion of the envelope of a wave packet. In isotropic materials, the group velocity is solely determined by the dispersion of the refractive index. However, in anisotropic materials such as birefringent crystals, the propagation of light depends strongly on the polarization state relative to the optical axis of the medium, the group velocity also consider the polarization state. Birefringence leads to the existence of two distinct polarization modes known as the ordinary and extraordinary waves. These modes experience different refractive indices and therefore propagate with different phase and group velocities. Such polarization-dependent group velocity plays a crucial role in nonlinear optics, ultrafast optics, and polarization control devices \cite{saleh2019fundamentals,yariv2007photonics,boyd2008nonlinear}.

In quantum optics \cite{scully1997quantum,mandel1996optical}, light must be treated as a quantized electromagnetic field. Photons generated from sources such as spontaneous parametric down-conversion (SPDC) \cite{kwiat1995new}are described as wave packets with finite spectral distributions. When such photons propagate through dispersive media, their temporal structure evolves according to the frequency dependence of the refractive index. In birefringent media, this evolution becomes polarization dependent, resulting in temporal walk-off between orthogonally polarized photon components \cite{vanselow2019ultra,couteau2018spontaneous}.

Nematic liquid crystals (NLCs) are attractive as controllable birefringent media because their optical properties can be modified by an external electric field. The elongated molecules in the nematic phase align along a preferred direction called the director. Application of a voltage across the liquid crystal cell causes a reorientation of the director, thereby modifying the effective refractive index experienced by extraordinary polarized light\cite{de1993physics,khoo2022liquid}.

The propagation of light in birefringent media and its impact on group velocity has been extensively studied in both classical\cite{wang2021group,wang2021biaxial,halvorsen2008superluminal} and quantum optics\cite{specht2009phase,smith2007photon,averchenko2024effect,petrov2004zero}. Previous works have primarily focused on classical descriptions of light propagation, where voltage-dependent birefringence is used to control phase retardation and intensity modulation\cite{kanseri2019broadband,kanseri2023degree,kanseri2024tunability}. In the context of quantum optics, studies have explored the role of birefringent media in manipulating polarization states, introducing temporal walk-off, and affecting entanglement properties of photon pairs\cite{Gya2026}. However, most existing approaches treat dispersion and voltage control separately, or consider simplified models that neglect the combined frequency and voltage dependence of the refractive index. Furthermore, the quantum description of photon wave packets in such dynamically tunable media remains relatively underexplored. 

In this work we develop a theoretical framework describing photon propagation and group velocity in voltage-controlled NLCs. The model incorporates both material dispersion and electrically tunable birefringence, allowing the refractive indices to depend simultaneously on frequency and applied voltage. By treating photons as quantum wave packets with finite spectral bandwidth, we analyze how different frequency components acquire distinct phase shifts and group delays during propagation. The voltage dependence is introduced through the reorientation of the liquid crystal director, which modifies the effective extraordinary refractive index and enables dynamic control of photon group velocity. This framework provides a unified description of temporal delay, phase evolution, and polarization entanglment manipulation .

\section{Classical description of group velocity in Birefringent Media}
The propagation of electromagnetic waves in a dielectric medium is governed by the dispersion relation, which relates the wavevector to the frequency of the wave. For a linear, non-magnetic medium, this relation can be expressed as \cite{brillouin2013wave}
\begin{equation}
k = \frac{n(\omega)\omega}{c},
\end{equation}
where $k$ is the magnitude of the wavevector, $\omega$ is the angular frequency, $c$ is the speed of light in vacuum, and $n(\omega)$ is the frequency-dependent refractive index of the medium. When considering the propagation of wave packets rather than monochromatic waves, the relevant quantity is the group velocity. The group velocity represents the speed at which the envelope of the wave packet propagates through the medium and is defined as
\begin{equation}
v_g = \frac{d\omega}{dk}.
\end{equation}
This definition can be rewritten in a more convenient form by inverting the derivative, which yields
\begin{equation}
v_g = \left(\frac{dk}{d\omega}\right)^{-1}.
\end{equation}
To evaluate this expression, we substitute the dispersion relation $k = n(\omega)\omega/c$ and differentiate with respect to frequency. This leads to
\begin{equation}
\frac{dk}{d\omega}=\frac{1}{c}\left[n(\omega)+\omega\frac{dn(\omega)}{d\omega}\right].
\end{equation}
Substituting this result back into the definition of group velocity provides an explicit expression for the group velocity in a dispersive medium
\begin{equation}
v_g=\frac{c}{n(\omega)+\omega\frac{dn(\omega)}{d\omega}}.
\end{equation}
The denominator of this expression is commonly referred to as the group refractive index, which incorporates both the refractive index and its frequency dependence. It is defined as
\begin{equation}
n_g = n(\omega) + \omega \frac{dn(\omega)}{d\omega}.
\end{equation}
In anisotropic media such as birefringent crystals or NLCs, the optical response is characterized by two principal refractive indices: the ordinary refractive index $n_o$, which is independent of propagation direction, and the extraordinary refractive index $n_e$, which varies with the orientation of the electric field relative to the optical axis. Consequently, two distinct group velocities, $v_{g,o}$ and $v_{g,e}$, arise in accordance with Equation (5) which we would discuss in further sections.
\section{Quantum description of photon wave packets}

In quantum optics the electromagnetic field is quantized and light is described in terms of photon creation and annihilation operators. A single photon is not necessarily monochromatic but can exist as a wave packet composed of a superposition of frequency components. The quantum state of such a photon wave packet can be written as \cite{smith2007photon,Gya2026}

\begin{equation}
|\psi\rangle =
\int d\omega \, f(\omega)a^\dagger(\omega)|0\rangle,
\end{equation}
where $a^\dagger(\omega)$ is the creation operator that creates a photon of frequency $\omega$, $|0\rangle$ represents the vacuum state of the electromagnetic field, and $f(\omega)$ is the spectral amplitude describing the frequency distribution of the photon wave packet. The spectral amplitude must satisfy a normalization condition so that the total probability of detecting the photon is unity. This requirement leads to the normalization relation

\begin{equation}
\int |f(\omega)|^2 d\omega =1.
\end{equation}
When the photon propagates through a dispersive medium of length $L$, each spectral component accumulates a phase that depends on the propagation constant of the medium. The phase acquired by a frequency component $\omega$ can therefore be written as
\begin{equation}
\phi(\omega)=k(\omega)L.
\end{equation}
As a result of this frequency-dependent phase shift, the quantum state of the photon after passing through the birefringent medium is modified. The output quantum state becomes
\begin{equation}
|\psi_{out}\rangle =
\int d\omega f(\omega)e^{ik(\omega)L}a^\dagger(\omega)|0\rangle.
\end{equation}
To understand how dispersion affects the propagation of the photon wave packet, it is useful to expand the wavevector $k(\omega)$ around the central frequency $\omega_0$ of the photon spectrum using a Taylor series expansion. This expansion can be expressed as
\begin{equation}
k(\omega)=k_0+k'_0(\omega-\omega_0)+\frac{1}{2}k''_0(\omega-\omega_0)^2+...
\end{equation}
where $k_0 = k(\omega_0)$ is the propagation constant at the central frequency, $k'_0$ represents the first derivative of the wavevector with respect to frequency evaluated at $\omega_0$, and $k''_0$ corresponds to the second derivative that accounts for group velocity dispersion. The first-order term in this expansion determines the delay experienced by the center of the photon wave packet as it travels through the medium. This propagation delay, known as the group delay, is given by
\begin{equation}
\tau = k'_0 L.
\end{equation}
The parameter $k'_0$ is directly related to the group velocity of the photon wave packet. The group velocity through the birefringent medium is defined as
\begin{equation}
v_g=\frac{1}{k'_0}.
\end{equation}
Therefore, the arrival time of the photon wave packet at the output of the medium depends directly on the group velocity determined by the dispersion properties of the material.
\section{Voltage-controlled photon group velocity and temporal delay in NLCs}
NLCs are birefringent materials composed of elongated, rod-shaped molecules that tend to align along a preferred direction known as the \textit{director}. Because of this molecular alignment, the optical response of the material depends on the polarization direction of the incident light. As a result, two distinct refractive indices exist corresponding to different polarization orientations. The refractive index experienced by light polarized perpendicular to the director is known as the ordinary refractive index and is denoted by $n_o$ whereas the refractive index experienced by light polarized parallel to the director is called the extraordinary refractive index and is denoted by $n_e$\cite{khoo2022liquid,de1993physics, kanseri2019broadband}.
When light propagates at an angle with respect to the director axis, the extraordinary polarization does not experience a constant refractive index. Instead, it experiences an effective refractive index that depends on the angle between the propagation direction and the director. This effective extraordinary refractive index can be expressed as

\begin{equation}
n_e^{eff}(\omega, \theta)=
\frac{n_o(\omega) n_e(\omega)}
{\sqrt{n_o^2(\omega)\sin^2\theta+n_e^2(\omega)\cos^2\theta}},
\end{equation}
where $\theta$ represents the angle between the propagation direction of the light and the director axis of the liquid crystal molecules. This angular dependence is responsible for the change in birefringence of the medium.

One of the most useful properties of NLCs is that the orientation of the director can be modified by applying an external electric field. When a voltage is applied across the liquid crystal cell, the molecules tend to rotate and align in the direction of electric field. As a consequence, the orientation angle becomes dependent on the applied voltage i.e. $\theta = \theta(V)$ and can be written as \cite{ding2019spectral,kanseri2019broadband,kanseri2023degree}
\begin{equation}
\theta(V) =
\begin{cases}
0, & V \leq V_{th} \\
\frac{\pi}{2} - 2 \tan^{-1} \left[ \exp\left(-\frac{V - V_{th}}{V_0}\right) \right], & V > V_{th}
\end{cases}
\end{equation}
where $V$ is the applied voltage, $V_{th}$ is the threshold voltage at which the director starts tilting and $V_0$ is a constant voltage and  $\theta$ ranges from $0$ to $\pi/2$. Because the effective extraordinary refractive index depends on the orientation angle, the application of voltage indirectly modifies the refractive index experienced by the extraordinary polarization. Therefore, the effective refractive index becomes a function of the applied voltage. The effective extraordinary refractive index is then calculated using the standard birefringence relation
\begin{equation}
n_e^{eff}(\omega, V) = \frac{n_o(\omega) n_e(\omega)}{\sqrt{n_o^2(\omega)\sin^2\theta(V) + n_e^2(\omega)\cos^2\theta(V)}},
\end{equation}
which captures the dependence of the refractive index on both frequency and molecular orientation.
In order to incorporate dispersion, the ordinary and extraordinary refractive indices are modeled as weakly frequency-dependent quantities using a linear approximation around a central frequency $\omega_0$. Specifically, the refractive indices are written as
\begin{equation}
n_o(\omega) = n_{o0} + b(\omega - \omega_0),
\end{equation}
\begin{equation}
n_e(\omega) = n_{e0} + a(\omega - \omega_0),
\end{equation}
where $n_{o0} = 1.5$ and $n_{e0} = 1.7$ are typical values for NLCs. The parameters $a$ and $b$, of the order of $10^{-16}$, represent small dispersion coefficients and ensure that the refractive index varies slowly with frequency, which is consistent with narrowband photon sources used in quantum optics experiments.

When quantum light propagates through a LC medium, the propagation constant depends on both the frequency of the photon and the refractive index of the material. For photons propagating in the extraordinary polarization mode inside the liquid crystal, the propagation constant can be written as

\begin{equation}
k_e(\omega,V)=\frac{\omega}{c}n_e^{eff}(\omega,V),
\end{equation}
where $\omega$ is the photon angular frequency, $c$ is the speed of light in vacuum, and $n_e^{eff}$ is the voltage-dependent effective extraordinary refractive index. To obtain an explicit expression for the group velocity, we differentiate the propagation constant with respect to frequency. This gives

\begin{equation}
\frac{dk_e(\omega, V)}{d\omega}=
\frac{1}{c}\left[n_e^{eff}(\omega, V)+\omega\frac{dn_e^{eff}(\omega, V)}{d\omega}\right].
\end{equation}
The group velocity for extraordinary mode can be written as
\begin{equation}
v_{g,e}(\omega,V)=
\frac{c}{n_e^{eff}(\omega,V)+\omega\frac{dn_e^{eff}(\omega,V)}{d\omega}}.
\end{equation}
This expression shows that the group velocity for extraordinary mode depends on both the refractive index and its frequency dispersion. Because the extraordinary refractive index depends on the applied voltage, the group velocity of the photon wave packet can be dynamically controlled.
For comparison, the group velocity associated with the ordinary polarization mode can be obtained using the ordinary refractive index. Since the ordinary index does not depend on the director orientation, the corresponding group velocity can be written as
\begin{equation}
v_{g,o}(\omega)=
\frac{c}{n_o(\omega)+\omega\frac{dn_o(\omega)}{d\omega}}.
\end{equation}
The difference between the ordinary and extraordinary refractive indices results in distinct propagation speeds and hence different group velocities for orthogonally polarized photon modes, leading to a temporal separation between these components that plays a crucial role in polarization-based quantum optical experiments.  As a consequence, the two components accumulate different propagation delays. The propagation time for a photon traveling through a medium of length $L$ for the extraordinary polarization mode, the propagation time can be written as
\begin{equation}
t_e(\omega,V)=\frac{L}{v_{g,e}(\omega,V)}.
\end{equation}
Here $v_{g,e}(V)$ is the voltage-dependent group velocity of the extraordinary photon mode inside the NLCs. Similarly, the propagation time for the ordinary polarization mode is given by
\begin{equation}
t_o(\omega)=\frac{L}{v_{g,o}(\omega)}.
\end{equation}
Since the ordinary refractive index does not dependent on the director orientation, the ordinary group velocity remains nearly constant with applied voltage. The temporal separation between the two polarization components is therefore determined by the difference between these propagation times. The resulting polarization-dependent time delay can be expressed as
\begin{equation}
\Delta t (\omega,V)= t_e (\omega, V)- t_o(\omega).
\end{equation}
Substituting the expressions for the propagation times yields
\begin{equation}
\Delta t (\omega,V) =L\left(\frac{1}{v_{g,e}(\omega,V)}-\frac{1}{v_{g,o}(\omega)}\right).
\end{equation}
substituting the previously derived expressions for the group velocities from equations (21-22) in equation (29), the delay can be written as
\begin{equation}
\begin{aligned}
\Delta t(\omega,V) = \frac{L}{c} \Big[ 
& n_e^{eff}(\omega,V) + \omega \frac{d n_e^{eff}(\omega,V)}{d\omega} \\
& - n_o(\omega) - \omega \frac{d n_o(\omega)}{d\omega}.
\Big]
\end{aligned}
\end{equation}
This is the expression for temporal delay which comes out to be the function of frequency and voltage.

\section{Two-photon entangled states}
Polarization-entangled photon pairs can be generated using spontaneous parametric down-conversion (SPDC) in nonlinear crystals, where a pump photon splits into signal and idler photons. Under suitable phase-matching conditions, a maximally entangled Bell state is produced as
\begin{equation}
|\Psi\rangle =
\frac{1}{\sqrt{2}}
(|H_sV_i\rangle + |V_sH_i\rangle),
\end{equation}
where $H$ and $V$ denote horizontal and vertical polarization states, while $s$ and $i$ label the signal and idler photons. This state represents a coherent superposition with perfect polarization correlations. When the entangled photons propagate through a birefringent medium such as a NLCs, the orthogonal polarization components experience different group velocities. This results in a temporal delay, which introduces a relative phase shift between the two components of the entangled state. The output state can therefore be written as
\begin{equation}
|\Psi_{out}\rangle =
\frac{1}{\sqrt{2}}
(|H_sV_i\rangle + e^{i\phi(V)}|V_sH_i\rangle),
\end{equation}
where the phase shift is given by
\begin{equation}
\phi(V)=\omega \Delta t(V),
\end{equation}
with $\omega$ being the angular frequency and $\Delta t(V)$ the voltage-dependent temporal delay. The corresponding density matrix reveals that the coherence terms depend explicitly on $\phi(V)$, indicating that the degree of quantum interference is controlled by the applied voltage. The interference visibility $\mathcal{V}=\cos(\phi)$ \cite{biswas2017interferometric,qureshi2019coherence,sun2021quantifying} can be expressed as a function of voltage as
\begin{equation}
\mathcal{V}(V) = |\cos(\phi(V))|,
\end{equation}
which exhibits oscillatory behavior as the phase varies with voltage. The strength of quantum correlations can be quantified using the Bell parameter $S = 2\sqrt{2}\mathcal{V}$\cite{qian2018temporal} in the CHSH inequality\cite{shin2019bell} and can be expressed in terms of voltage,
\begin{equation}
S(V) = 2\sqrt{2}\,|\cos(\phi(V))|.
\end{equation}
Violation of Bell’s inequality ($S > 2$) confirms the presence of entanglement, while its voltage dependence enables controlled tuning between quantum and classical regimes.

\section{Results and discusssion}
\begin{figure}[ht]
\centering
\includegraphics[width=1\linewidth]{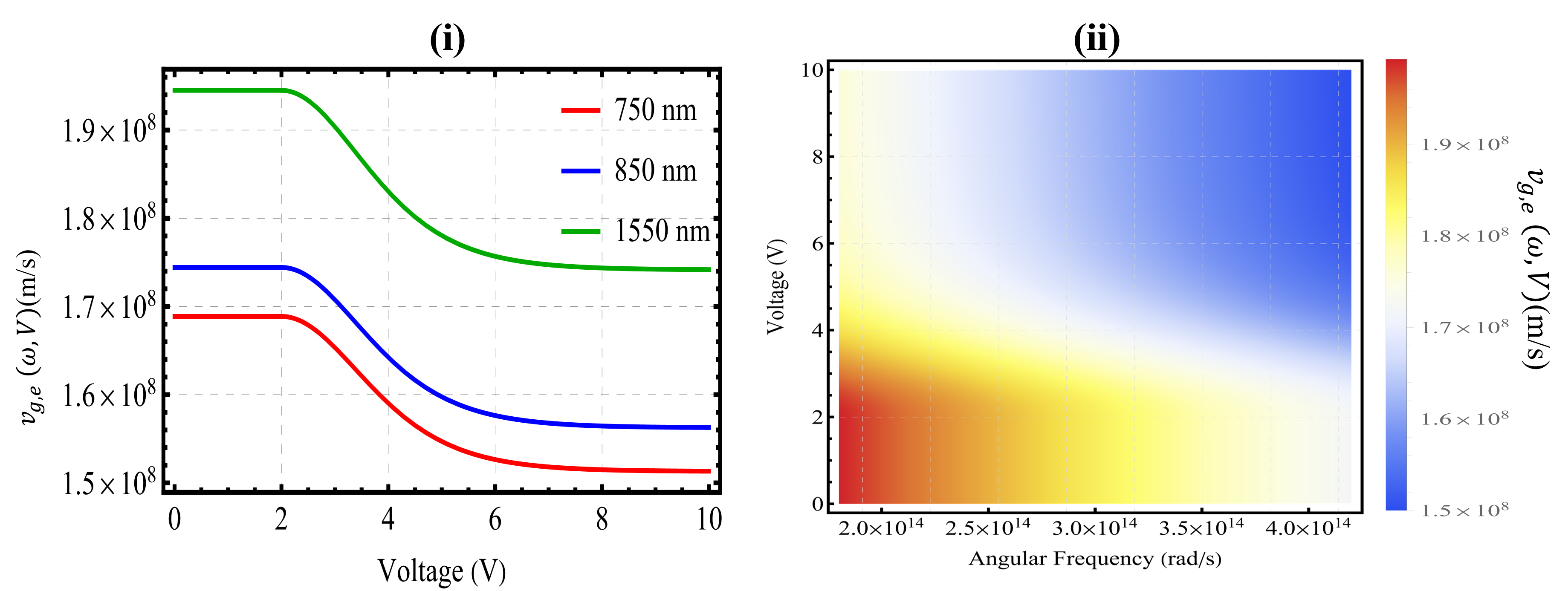}
\caption{(i)Group velocity as a function of applied voltage for three different wavelengths (750 nm, 850 nm, and 1550 nm) in a nematic liquid crystal. (ii)Heat map showing the variation of photon group velocity $v_g(\omega, V)$ as a function of angular frequency $\omega$ and applied voltage $V$ . The color scale represents the magnitude of the group velocity, illustrating the combined effects of material dispersion and voltage-controlled birefringence.}
\label{Fig1}
\end{figure}

The values from equation (16), equation (17) and equation (18) can be substituted in equaion (21) and the group velocity for extraordinary, using this equation, is plotted in Figure \ref{Fig1}. Figure \ref{Fig1}(i) shows the variation of photon group velocity as a function of applied voltage for three different optical frequencies corresponding to wavelengths of approximately 750 nm, 850 nm, and 1550 nm.  
The resulting curves demonstrate that the group velocity varies continuously with applied voltage for all three frequencies. At the same time, a clear separation between the curves corresponding to different wavelengths is observed, which arises due to dispersion. The threshold voltage in practical LC cells depends on several parameters; however, for the present theoretical analysis, a threshold voltage value of 2 V is assumed\cite{jiao2008alignment,nie2007anchoring}. At low voltages ($V \leq V_{th}$), the curves remain nearly constant. In this regime, the liquid crystal director is not significantly reoriented, and hence the birefringence-induced phase shift remains unchanged. Higher frequencies (shorter wavelengths) experience a slightly stronger influence of the dispersive term $\omega \frac{\partial n}{\partial \omega}$, leading to lower group velocities compared to longer wavelengths such as 1550 nm. This indicates that different spectral components of a photon wave packet will propagate at different speeds, resulting in frequency-dependent temporal delays. The figure demonstrates that NLCs function as electrically tunable dispersive media in which both the magnitude of the group velocity and its spectral dependence can be controlled. This control mechanism has important implications for quantum optics, as it enables precise manipulation of photon arrival times. Such control is essential for applications in quantum communication, interferometry, and entanglement engineering, where even small variations in applied voltage can significantly affect quantum interference and coherence properties.
The heat map in Figure \ref{Fig1}(ii) is another kind of depiction of the photon group velocity $v_g(\omega, V)$ on both angular frequency $\omega$ and applied voltage $V$. The horizontal axis represents the angular frequency in the optical range, while the vertical axis corresponds to the applied voltage. The color scale indicates the magnitude of the group velocity, with gradual variations reflecting changes in the optical properties of the medium. The heat map clearly shows a continuous variation of group velocity along the voltage axis, confirming the electrically tunable nature of the medium. Along the frequency axis, a systematic variation in group velocity is observed due to material dispersion. Higher angular frequencies exhibit slightly lower group velocities compared to lower frequencies. This behavior arises from the frequency dependence of the refractive index and its derivative, which together determine the group refractive index. The smooth gradient observed in the heat map indicates that both voltage and frequency serve as independent control parameters for tuning photon propagation. The voltage-dependent variation enables dynamic control of photon arrival times, while the frequency dependence highlights the role of dispersion in shaping the temporal evolution of photon wave packets.

\begin{figure}[ht]
\centering
\includegraphics[width=1\linewidth]{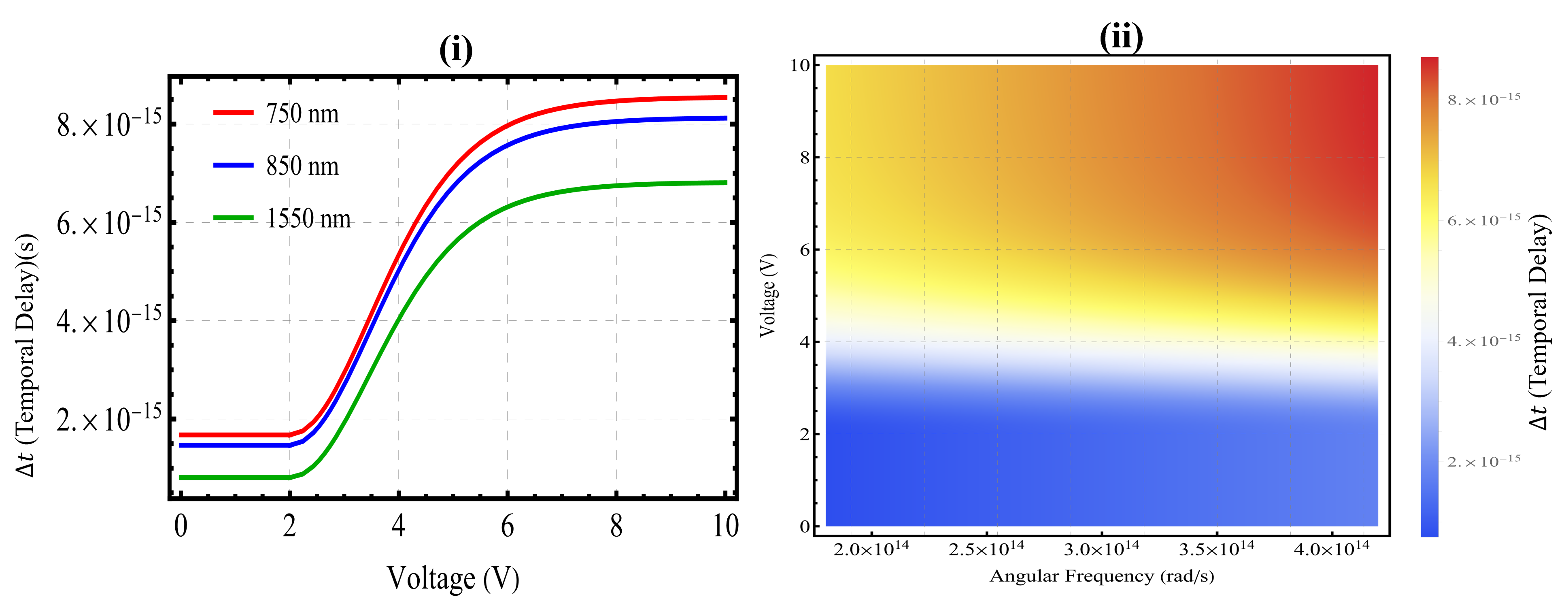}
\caption{(i)Temporal delay $\Delta t$ as a function of applied voltage for three different wavelengths (750 nm, 850 nm, and 1550 nm) in a nematic liquid crystal cell of thickness $L = 10\,\mu$m.(ii) The heat map shows the teporal delay in color scale, frequency on horizontal axis and voltage on vertical axis.}
\label{Fig2}
\end{figure}

The temporal delay $\Delta t(\omega, V)$, obtained using equation.~(27), is plotted as a function of the applied voltage for different wavelengths in Figure \ref{Fig2}. Each curve in the figure \ref{Fig2}(i) corresponds to a  wavelengths 750 nm, 850 nm, and 1550 nm, illustrating how the delay evolves as the voltage is varied from 0 to 10 V. A clear separation between the curves corresponding to different wavelengths is evident, highlighting the role of dispersion. Since the refractive indices are frequency dependent, the group velocity contains a dispersive contribution through the term $\omega \frac{dn}{d\omega}$. As a result, shorter wavelengths (such as 750 nm), which correspond to higher angular frequencies, experience a stronger influence of dispersion compared to longer wavelengths like 1550 nm. This leads to distinct group velocities and consequently different temporal delays for each wavelength. Physically, this implies that different spectral components of a photon wave packet propagate with slightly different arrival times, giving rise to spectral walk-off. Although the absolute magnitude of the temporal delay is small due to the micrometer-scale thickness of the LC cell, its impact in quantum optical systems is significant. This temporal separation between orthogonally polarized components is known as temporal walk-off~\cite{wang2003efficiency}. By adjusting the applied voltage, it is therefore possible to precisely control this walk-off, enabling tunable manipulation of polarization dynamics. Such voltage-controlled temporal delays are particularly important in polarization-based quantum optical experiments, where precise control over phase and temporal overlap is essential. The ability to dynamically tune these parameters makes NLCs a versatile platform for controlling photon wave packets, engineering quantum interference, and manipulating entangled states. The heatmap in Figure \ref{Fig2}(ii) is another visual depiction of temporal delay varies smoothly with both frequency and applied voltage. Along the voltage axis, the delay changes continuously due to the reorientation of the liquid crystal director, which alters the birefringence and hence the difference in propagation speeds of the two polarization modes. Along the frequency axis, the delay exhibits variation due to dispersion, indicating that different spectral components of a photon wave packet experience different propagation times. Because the liquid crystal thickness is in the micrometer range, the absolute magnitude of the delay is small. However, even such small delays are significant in quantum optical systems, since they correspond to phase shifts of the form $\phi(\omega, V) = \omega \Delta t(\omega, V)$, which can be large due to the high optical frequencies involved. This implies that voltage-controlled birefringence can be used to precisely manipulate phase relationships between polarization modes.

\begin{figure}[ht]
\centering
\includegraphics[width=.5\linewidth]{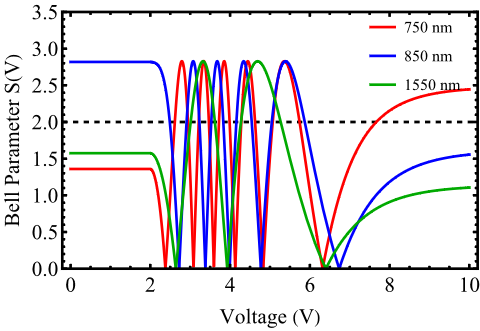}
\caption{Bell parameter $S(V)$ versus applied voltage for different wavelengths. The line $S=2$ marks the classical limit, while $S>2$ indicates Bell inequality violation. Voltage-dependent oscillations arise from birefringence-induced phase shifts, demonstrating electrical control of quantum entanglement.
}
\label{bellpara}
\end{figure}

From equation (33),The variation of the Bell parameter $S(V)$  as a function of the applied voltage for three previously considered  wavelengths is shown in the Figure \ref{bellpara}. The Bell parameter is a direct measure of quantum correlations and is used to test the violation of the Clauser–Horne–Shimony–Holt (CHSH) inequality. The horizontal line at $S = 2$ represents the classical limit, below which the correlations can be explained by local realistic theories, while values exceeding this threshold indicate the presence of quantum entanglement.
Up to the threshold voltage (i.e., 2 V), the Bell parameter remains constant. As the applied voltage exceeds the threshold value of voltage, a clear oscillatory behavior in $S(V)$ is observed for all three wavelengths. This behavior arises from the voltage-dependent phase shift $\phi(V) = \omega \Delta t(V)$, which is induced by the change in the effective extraordinary refractive index due to director reorientation. The modulation of $\phi(V)$ directly affects the quantum interference between the two components of the entangled state, resulting in periodic variations of the Bell parameter. A key feature of the figure is the presence of regions where $S(V) > 2$, indicating violation of Bell’s inequality and the existence of nonlocal quantum correlations. As the voltage is varied, the system transitions between regimes of strong entanglement ($S > 2$) and reduced or classical correlations ($S \leq 2$). This demonstrates that the degree of entanglement can be dynamically controlled through an external electrical parameter. The three curves corresponding to different wavelengths exhibit distinct oscillation patterns, highlighting the role of dispersion. Since the phase shift depends on angular frequency, shorter wavelengths (higher frequencies) accumulate phase more rapidly, leading to faster oscillations in the Bell parameter. In contrast, longer wavelengths such as 1550 nm show comparatively slower variations. This confirms that both dispersion and voltage jointly influence the evolution of quantum correlations. Overall, the figure clearly demonstrates that a NLC cell can function as an electrically tunable quantum device capable of controlling entanglement. By adjusting the applied voltage, it is possible to manipulate the phase, interference visibility, and Bell parameter of polarization-entangled photon pairs. This provides a practical mechanism for active control of the quantum states, which is highly relevant for applications in quantum communication, quantum information processing, and reconfigurable photonic systems.photonic devices.
\section{Conclusion}
In this work, we have developed a unified theoretical framework for describing the propagation of the quantum states of light in voltage-controlled NLCs, incorporating both material dispersion and electrically tunable birefringence within a single consistent model. Starting from the quantum description of photon wave packets, analytical expressions for group velocity, temporal walk-off, and phase evolution were derived, revealing how voltage-induced director reorientation enables dynamic control of polarization-dependent propagation. The results demonstrate that the extraordinary photon group velocity and the resulting temporal delay can be continuously tuned using an external electric field, leading to controllable phase shifts that directly influence quantum interference and entanglement. The novelty of this work lies in the simultaneous treatment of frequency-dependent dispersion, voltage-controlled anisotropy, and quantum wave packet dynamics, which has not been addressed in a unified manner in earlier studies. This approach provides deeper insight into photon dynamics in anisotropic media and establishes NLCs as versatile, electrically controllable platforms for applications such as quantum delay lines, phase shifters, and entanglement manipulation in quantum communication and photonic technologies.

%
%

\ack{The authors gratefully acknowledge Prof.\ Bhaskar Kanseri (IIT Delhi) for his valuable insights and guidance in the past, which were instrumental in shaping the research idea presented in this work.}





\bibliographystyle{iopart-num}
\bibliography{references}

\end{document}